\documentclass[fleqn,10pt]{wlscirep}
\usepackage[utf8]{inputenc}
\usepackage[T1]{fontenc}
\usepackage{siunitx} 
\usepackage{graphicx}
\usepackage{hyperref}
\usepackage{textcomp}
\usepackage[section]{placeins}
\usepackage{lineno}
\usepackage{soul}
\usepackage{xcolor}

\title{Spectral and spatial shaping of laser-driven proton beams using a pulsed high-field magnet beamline}

\author[1,2,+]{Florian-Emanuel Brack}
\author[1]{Florian Kroll}
\author[1,2]{Lennart Gaus}
\author[1,2]{Constantin Bernert}
\author[1,3]{Elke Beyreuther}
\author[1,2]{Thomas E. Cowan}
\author[1,3]{Leonhard Karsch}
\author[1]{Stephan Kraft}
\author[3]{Leoni A. Kunz-Schughart}
\author[1]{Elisabeth Lessmann}
\author[1]{Josefine Metzkes-Ng}
\author[1]{Lieselotte Obst-Huebl\footnote{Now at Lawrence Berkeley National Laboratory}}
\author[1,3]{Jörg Pawelke}
\author[1,2]{Martin Rehwald}
\author[1]{Hans-Peter Schlenvoigt}
\author[1,2]{Ulrich Schramm}
\author[1]{Manfred Sobiella}
\author[4]{Emília Rita Szabó}
\author[1,2]{Tim Ziegler}
\author[1]{Karl Zeil}

\affil[1]{Helmholtz-Zentrum Dresden -- Rossendorf, 01328 Dresden, Germany}
\affil[2]{Technische Universität Dresden, 01062 Dresden, Germany}
\affil[3]{OncoRay – National Center for Radiation Research in Oncology, Faculty of Medicine and University Hospital Carl Gustav Carus, TU Dresden and Helmholtz-Zentrum Dresden–Rossendorf, Dresden, Germany}
\affil[4]{Attosecond Light Pulse Source, ELI-HU Nonprofit Ltd., Szeged, Hungary}
\affil[+]{f.brack@hzdr.de}

\begin{abstract}
Intense laser-driven proton pulses, inherently broadband and highly divergent, pose a challenge to established beamline concepts on the path to application-adapted irradiation field formation, particularly for 3D. Here we experimentally show the successful implementation of a highly efficient (\SI{50}{\percent} transmission) and tuneable dual pulsed solenoid setup to generate a homogeneous (laterally and in depth) volumetric dose distribution (cylindrical volume of 5mm diameter and depth) at a single pulse dose of \SI{0.7}{\gray} via multi-energy slice selection from the broad input spectrum. The experiments were conducted at the Petawatt beam of the Dresden Laser Acceleration Source \textit{Draco} and were aided by a predictive simulation model verified by proton transport studies. With the characterised beamline we investigated manipulation and matching of lateral and depth dose profiles to various desired applications and targets. Using an adapted dose profile, we performed a first proof-of-technical-concept laser-driven proton irradiation of volumetric \textit{in-vitro} tumour tissue (SAS spheroids) to demonstrate concurrent operation of laser accelerator, beam shaping, dosimetry and irradiation procedure of volumetric biological samples.
\end{abstract}

\begin{document}
\flushbottom
\maketitle
\thispagestyle{empty}
\section*{Introduction} \label{sec:intro}

Laser plasma accelerators\cite{Daido2012, Macchi2013} can deliver intense and energetic proton bunches of sub-\si{\pico\second} duration and with unique characteristics\cite{Cowan2004} as compact sources. By that they have the potential for a wide range of multi-disciplinary applications. 
This includes warm dense matter research \cite{Patel2003}, probing of ultra-fast plasma dynamics\cite{Romagnani2005}, material research\cite{Dromey2016} and archaeological surveys\cite{Barberio2017}, injector sources for conventional accelerator structures\cite{Antici2011,Busold2014,Busold2015} or radiobiology studies of laser-driven proton and ion beams\cite{Yogo2009,Kraft2010,Yogo2011,Bin2012,Doria2012,Zeil2013a,Hanton2019,Bayart2019} as well as translational research in laser-driven radio-oncology\cite{Masood2014}. In general these applications require specific beam qualities, such as controlled spectral and spatial shapes, particle number as well as sufficient reproducibility and stability. In parallel to ongoing development and improvement of the high power laser sources on the petawatt (PW) level\cite{Danson2015,Schramm2017,Gales2018}, continuous efforts in the field have been undertaken to study and optimise the laser-matter interaction. Advanced acceleration schemes\cite{Qiao2019} and sophisticated targetry\cite{Buffechoux2010,Zeil2014,Obst2017,Prencipe2017,Hilz2018} are recognised as possible routes to improving key features of the laser accelerated proton beams, such as narrowed spectra or enhanced intensity and energy. However, target normal sheath acceleration (TNSA) from thin solid-density foils remains today the best established and most stable acceleration mechanism. Therefore, it is most commonly used for proof-of-concept experiments in the mentioned range of applications. 

In short pulse driven TNSA, protons originating from the target surface layers gain energy along the target normal direction due to space charge fields set up by fast electrons\cite{Zeil2012}, which in turn have been accelerated by the relativistic laser pulses at the front surface plasma\cite{Snavely2000}. Intrinsically, TNSA-accelerated proton pulses feature broad exponentially decreasing spectra with cut-off energies of tens of \si{\mega\electronvolt} up to approximately \SI{90}{\mega\electronvolt} \cite{Wagner2016,Higginson2018} (currently) and an energy-dependent half-opening angle of up to \SI{20}{\degree}. 
As a consequence, tailored transport and beam shaping techniques have to be used to prepare application specific beam parameters\cite{Nishiuchi2009,Busold2015,Romano2016,Pommarel2017,Zhu2019}. Ideally, innovative laser plasma based concepts\cite{Toncian2006,Kar2016,Obst-Huebl2018} might be exploited for initial beam manipulation. Capture, transport and focusing of the strongly divergent and broad bandwidth proton beams represent the most challenging task. In previous applications\cite{Bin2012,Bayart2019}, dedicated compact permanent quadrupole magnet assemblies\cite{Eichner2007,Ter-Avetisyan2008,Schollmeier2008,Bin2012} were successfully applied, but may experience limitations in transmission efficiency due to the asymmetric focusing / defocusing characteristics in the transverse planes. The implementation of large aperture high-field solenoids, providing symmetric focusing conditions, has thus been discussed to circumvent this problem\cite{Burris-Mog2011,Busold2014a,Jahn2018}.While field strengths of permanent magnets or direct current electromagnets are typically limited by saturation of the core materials to below two Tesla, non-destructive pulsed high-field magnets can provide up to several tens of Tesla. This tremendously reduces size and weight of the beamline structures\cite{Masood2017}. Driven by pulsed power supplies the magnetic field can be tuned independently per pulse.

Special challenges arise for the application of laser-driven proton beamlines as dose delivery systems for radiobiological studies, in particular if three dimensional volumetric biological samples are envisaged. 
There, homogeneous dose distribution over the sample volume is mandatory. Longitudinal homogeneity over a certain depth can be obtained from a spread out Bragg peak (SOBP) which requires superimposed protons of a correspondingly broad energy window with weighted spectral intensity.
Equally important is the lateral dose homogeneity, which is necessary for an evenly distributed absorbed dose throughout the entire sample volume. A sufficiently high dose rate of the order of $\gtrsim\SI{1}{\gray\per\minute}$, appropriate shielding of the sample against secondary radiation, and real-time dose control for the radiobiological sample, that has to be irradiated in-air, conclude the requirements\cite{Zeil2013a}. 

In the following work we present the design and optimisation of a compact laser-driven proton beamline based on two pulsed high-field solenoid lenses and its implementation at the \textit{Draco} laser facility for dose-controlled irradiation studies of three-dimensional biological samples. This appears in the context of an extensive translational research programme focusing on radiobiological \textit{in-vivo} studies\cite{Bruchner2014,Oppelt2015,Beyreuther2017} via irradiation of 3D tumour entities with low-energy high-dose-rate proton bunches. With the presented beamline the generation of volumetrically homogeneous SOPB dose distributions in a single shot is demonstrated for target volumes of up to  \SI[product-units=power]{5x5x5}{\milli\metre} to be irradiated with a dose of about \SI{1}{\gray} per shot. The SOBP is produced by mixing multiple proton energy contributions in a single shot, similar to the concept proposed by Masood et al.\cite{Masood2017}, and therefore taking full advantage of the broad energy spectrum inherent to the TNSA mechanism. 


\section*{Concept and setup of a laser-driven proton beamline at Draco} \label{sec:sec1}

\begin{figure}[tbph]
\centering
\includegraphics[width=0.95\linewidth]{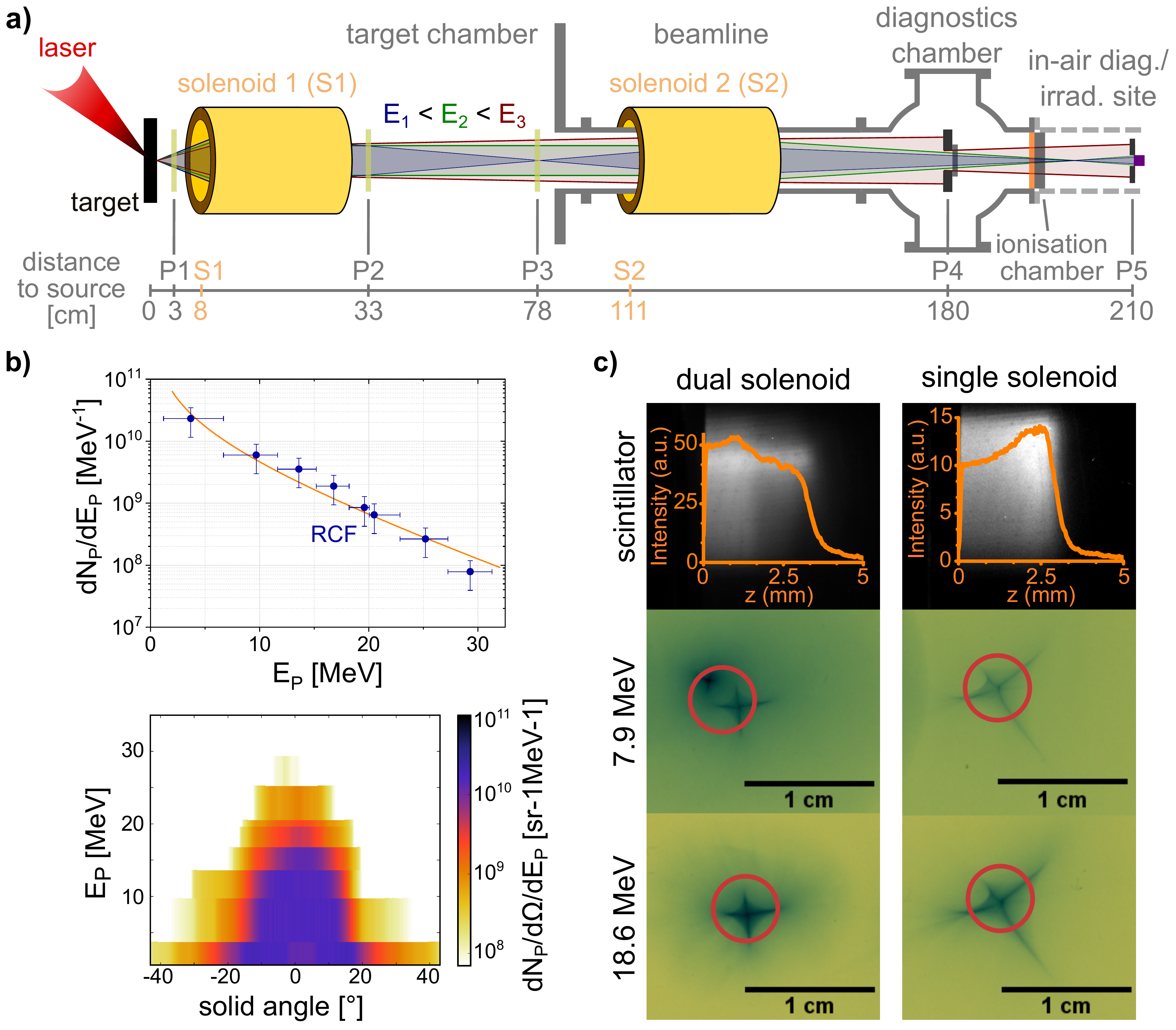}
\caption{
\textbf{a)} Schematic of the proton beamline at the \textit{Draco} laser facility. At positions P\SIrange[range-phrase=--]{1}{5}{} detectors can be installed. \textbf{b)} Representative proton source characteristics from RCF stack measurements: integrated TNSA proton spectrum (top) and the angular distribution (bottom) for full energy \textit{Draco}~PW shot on a \SI{80}{\nano\metre} plastic target. 
The orange line represents a parametrisation to the shown RCF data. 
\textbf{c)} 
Penetration depth (bulk scintillator, top) and lateral dose distributions of proton beams of main energy $\sim\SI{19}{\MeV}$ focused at P4 via single solenoid transport (right column) or dual solenoid transport (left column). The lateral dose distributions are recorded on RCF (corresponding Bragg peak energies \SI{7.9}{\MeV} and \SI{18.6}{\MeV}) and the red circles represent a typical aperture size (\SI{5}{\milli\metre} diameter) for proposed irradiation experiments.}
\label{fig:setup}
\end{figure}

The presented beamline is installed at one of the target areas of the \textit{Draco} laser facility at Helmholtz-Zentrum Dresden--Rossendorf (HZDR)\cite{Schramm2017}. Its main design features are presented in fig.~\ref{fig:setup}a). Using the Petawatt beam of \textit{Draco}\cite{Zeil2010,Obst2018} ($E_L=\SI{13}{\joule}$ after recollimating single-pass plasma mirror, $\tau = \SI{30}{\femto\second}$, \SI{3}{\micro\metre} FWHM spot size) on \SIrange{80}{200}{\nano\metre} plastic targets, we accelerate protons via TNSA which are then transported by the key components of the beamline: two identically designed pulsed high-field solenoids - one in close vicinity to the laser target installed in vacuum (solenoid S1) and one outside of the chamber (solenoid S2, technical details given in the methods section). Further downstream is a diagnostic chamber equipped with a thin transmission ionisation chamber for online dose monitoring, followed by a \SI{25}{\micro\metre} Kapton window acting as the vacuum-air boundary. The irradiation site is located at the end of the beamline, where either radiobiological samples or in-air diagnostics can be installed and tested\cite{Haffa2019}. At positions P\SIrange[range-phrase=--]{1}{5}{}, detectors (stacks of self developing radiochromic films (RCF), scintillator blocks, ultra-fast diamond detector) or beam-manipulating elements (apertures, scatter foils) can be introduced. The following paragraphs explain the conceptual ideas behind the beamline setup for radiobiological \textit{in-vivo} studies on three-dimensional tumour entities with laser-driven protons.

Radiobiological studies on volumetric samples generally require a homogeneous dose distribution throughout the entire sample. Generating such a dose distribution from a TNSA proton source requires spectral and spatial modification of the divergent beam. 
In order to maintain a high throughput, solenoid S1 with a \SI{40}{\milli\metre} bore opening diameter is placed \SI{8}{\centi\metre} behind the laser target, resulting in a geometrical acceptance angle of \SI{14}{\degree} (half-angle). 
S1 is used to efficiently capture the broad spectrum emitted by the laser-driven source. Up to an energy class $E_2$, defined by the variable solenoid setting, protons can be focused to a real focus downstream, for energies larger than $E_2$ the initial divergence is only reduced. To conveniently distinguish, we from now on refer to the beam of energy $E_2$ as collimated (ignoring the influence of finite emittance and energy spread).
This collimated beam propagates in vacuum towards solenoid S2. S2 is set to focus protons of energy $E_2$ in front of P5 to generate an expanded beam at sample position. For irradiation studies where lateral dose homogenisation is required, a scatter foil is installed at P4 as well as an energy-selecting aperture, which suppresses unwanted energies. Lastly, the irradiation field size at P5 is defined by a proton beam block with an aperture according to the sample geometry. 

Solenoid magnets are chromatic focusing devices\cite{Kumar2009} with the focal length $f$ being proportional to the particle momentum $f \propto p(E)^2$. With fig.~\ref{fig:setup}a) we present a specific energy class $E_1 < E_2$ (blue beam) that, for a given setting of S1, is focused between S1 and S2 in a way that this fraction of the beam is efficiently recaptured by S2 and finally focused by S2 to the same position as protons with $E_2$ (green beam) in the combined system. This leads to two individual and thus tuneable fractions of the broad TNSA spectrum being superimposed at the plane of interest.
Fig.~\ref{fig:setup}c), top left, shows the scintillation light the depth dose distribution in a scintillator induces at P4 (no scatter foil or aperture used) where the two transported spectral components are clearly visible as two distinguishable penetration depths. The RCF images below show corresponding lateral focal spot shapes. RCF data were obtained by stacking films, where protons reach a film in a certain depth according to their Bragg-peak energy. Two focal spots are visible on the \SI{7.9}{\MeV} RCF, but only one remains visible on the \SI{18.6}{\MeV} RCF. The sensitivity of the solenoid alignment allowed us to spatially separate the different transported spectral parts without measurable impact on lateral focal spot shape. The red circles on the RCFs depict the sample size (\SI{5}{\milli\metre} diameter), showing that a circular focal spot is preferred in order to minimise proton loss due to the tails of a star-like focus. 
Nevertheless, for a given sample size, a deliberately introduced ``misalignment'' of the solenoids in combination with an aperture, e.g. at P4, enables the control of the ratio of the focal intensities of both spectral components in the target plane. For clarification, imagine the red circle in fig.~\ref{fig:setup}c) representing this aperture. By moving the aperture (or by changing the overall pointing of the transported beam) the transmission can be tailored, ranging from the scenario where one energy component is fully blocked while the other is completely transmitted to an equalisation of both transmission efficiencies.
    
Such adjustments cannot be done in the single solenoid transport setting using only S1 (shown in fig.~\ref{fig:setup}c), right), as only one energy band is focused at the detector plane P4. Here, the scintillator shows a depth dose distribution corresponding to a quasi-monoenergetic beam (FWHM energy spread about \SIrange{10}{15}{\percent}), while with the RCFs a similar, yet larger focal shape is detected. 
This star-like focal spot shape has been observed at several laser facilities using different focusing solenoids\cite{Busold2014a,Jahn2018,Kroll2018}. Its non-trivial cause is currently under investigation using both laser- and conventional accelerators.

Due to the flexibility of the beamline with its different setups, it is suited for a broad range of applications. Notable is the interaction of the TNSA source and the dual solenoid setup which enables the transport of two separate spectral components on a single-shot basis.


\section*{Beamline modelling and experimental verification}

To predict solenoid parameters for optimised beam transport, we developed a simulation model of the beamline and its components using General Particle Tracer, a 3D particle tracing software. Each solenoid is handcrafted leading to small deviations in the complex winding geometry (winding steepness, exact distances between winding layers, in- and outlet of the wire, etc.) unknown after completion. Therefore, a reproduction of the solenoids' complex internal structure in simulations is of limited precision. Hence, the simulations use current loops arranged in layers according to the actual used winding and layer numbers for predicting the solenoid field maps. 

We characterised the beamline experimentally and adapted the model accordingly.
Keeping the simulated winding geometry constant, the peak solenoid current $I_\mathrm{S}$ was chosen as the optimisation parameter. 
Our aim is to find the translation factor $\alpha$. This factor is supposed to predict optimal experiment parameters from simulation studies by translating the solenoid current $I_\mathrm{S,sim}$ found in simulation to the according peak solenoid current $I_\mathrm{S,exp}$, measured during experiment, following $I_\mathrm{S,exp}=\alpha \cdot I_\mathrm{S,sim}$.

\begin{figure}[tbp]
\centering
\includegraphics[width=0.95\linewidth]{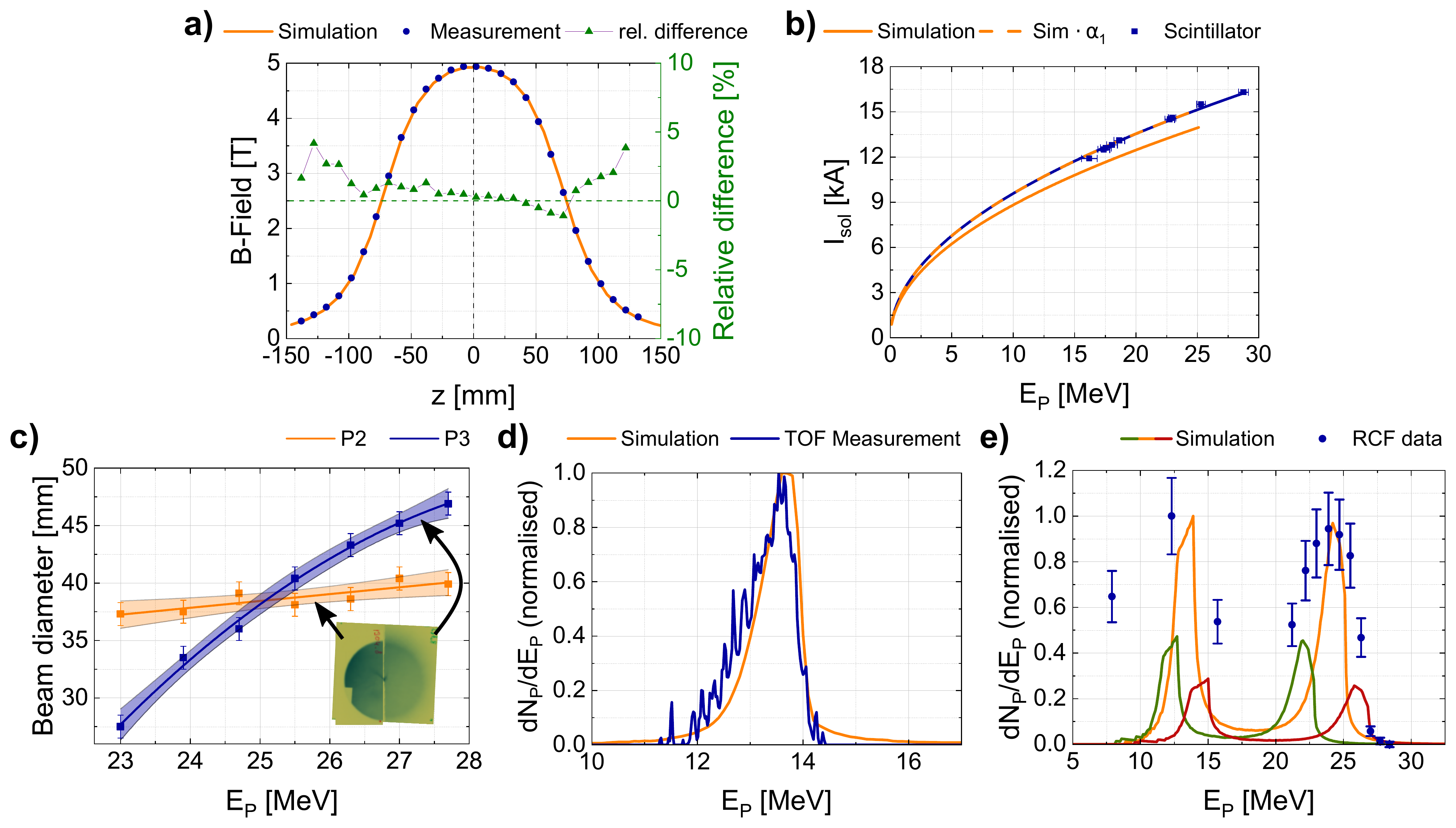}
\caption{Experimental verification of the beamline model and determination of translation factors $\alpha_{0-3}$ and $\beta$: \textbf{a)} Comparison of measured and simulated (GPT) $B$-field on axis of S1 resulting in $\alpha_0 = 1.06$. The zero position corresponds to the solenoid centre. \textbf{b)} Relation between focused proton energy (at P4) and applied solenoid current for simulations (orange) and experiment (blue). The dashed orange line shows the simulated values multiplied with the translation factor $\alpha_1 = 1.08$. The experimentally focused energy was determined from the penetration depth of the protons in a scintillator block at P4. \textbf{c)} Energy resolved beam size formation for free propagation to P2 (orange) and P3 (blue) leading to $\alpha_2 = 1.05$. Coloured areas represent \SI{95}{\percent} confidence band of the fit functions. Of the two shown RCFs (corresponding Bragg peak energy \SI{25,5}{\mega\electronvolt}) in the inset, the left RCF (placed at P2) has half the size of the right RCF (at P3) and was blocking/detecting half of the beam. \textbf{d)} Normalised proton energy spectrum from time-of-flight (blue) in comparison to simulation (orange) for $\alpha_3=1.14$. \textbf{e)} Comparison of a normalised experimental transmission spectrum (blue dots, experimental data originating from RCF stack measurement at P4) for dual solenoid transport with equivalent simulation using $\alpha = \beta = 1.09$ (orange line). The higher discrepancy for lower energetic protons is due to their larger divergence angle and the fact that simulation particles are distributed homogeneously over the corresponding angle whereas TNSA protons exhibit a Gaussian-like angular distribution.
The shown simulations in green ($\alpha = \beta = 1.14$) and red ($\alpha = \beta = 1.05$) indicate the sensitivity of our model. Both graphs are normalised with respect to the transmitted proton number for $\alpha = \beta = 1.09$.
}
\label{fig:1sol}
\end{figure}

An initial translation factor $\alpha_0$ was determined by comparing the simulated and measured magnetic field strength along the main solenoid axis. A measurement of a complete 3D field map with high resolution, including all fringe fields, standard for permanent magnets or DC devices, is not practical for pulsed high-field solenoids at relatively low repetition rate. The experimental data was acquired via a Hall-probe suitable for measurements in pulsed high magnetic fields\cite{Mironov2010a}. 
Fig.~\ref{fig:1sol}a) shows the comparison of the $B$-field measurement (50 pulses) to the simulated $B$-field on axis resulting in a translation factor of $\alpha_0 = \SI{1.06}{}$, adjusted to minimise the difference of the peak field strengths. 
The simulated field distribution is in particularly good agreement with the measurement, allowing us to perform particle tracing studies using GPT. 

The derivation of $\alpha_0$ only takes a small fraction of the $B$-field map into account. Yet, charged particle motion inside a solenoid is also strongly affected by fringe field shape and amplitude. To improve the applicability of $\alpha_0$ for predictive simulation, three independent, application oriented methods were studied at the \textit{Draco} laser facility, yielding three independent factors $\alpha_{1,2,3}$. All three employ single-shot diagnostics in consideration of the pulsed operation of the solenoids and TNSA source. 

The first method makes use of the correlation between solenoid focal length and magnetic field strength. Keeping the detector plane fixed at P4, we varied the solenoid current, and therefore the $B$-field, to focus protons of different energies onto a scintillator block. The maximum penetration depth of the focused protons in the scintillator (as seen in fig.~\ref{fig:setup}c)) corresponds to their kinetic energy. For a range of applied currents $I_\mathrm{S1}$, protons of different kinetic energy $E_\mathrm{P}$ were focused. The comparison to a set of simulations replicating the measurements (see fig.~\ref{fig:1sol}b)) using square root fits yields a translation factor $\alpha_1 = \SI[separate-uncertainty = true]{1.08+-0,02}{}$.

Fig.~\ref{fig:1sol}c) summarises experimental results, where we aimed to determine the kinetic energy of protons which are collimated by S1 at a certain fixed peak $B$-field. 
In order to do so, the solenoid peak current was fixed at $I_{S1} = \SI{14}{\kilo\ampere}$. An RCF stack at P2 covered half of the solenoid aperture and simultaneously a second RCF stack at P3 covered the full aperture. The graphs in fig.~\ref{fig:1sol}c) show the energy dependent beam size formation at both positions for freely propagating protons downstream of S1. Two irradiated films, corresponding to a Bragg peak energy of \SI{25.5}{\mega\electronvolt} are shown in fig.~\ref{fig:1sol}c) demonstrating a well collimated beam. The slopes of the proton beam diameters as a function of their kinetic energy are different for positions P2 and P3, because of the different propagation lengths in combination with the energy dependent focal lengths. The intersection of the drawn fit functions marks the collimated energy, i.e. $E_\mathrm{coll} = \SI[separate-uncertainty = true]{25.1+-0,4}{\mega\electronvolt}$. GPT-simulations were performed, using a divergent source of $\SI{25.1}{\MeV}$ protons. By altering the model current $I_\mathrm{sim} = \SI{14}{\kilo\ampere}/{\alpha_2}$, the mean divergence angle behind S1 was minimised. According to this method a translation factor $\alpha_2 = \SI[separate-uncertainty = true]{1.05+-0,02}{}$ was derived.

The third complementary method to determine the translation factor is the analysis of the spectral distribution of the proton beam via the time-of-flight (TOF) method\cite{Milluzzo2019}. A fast diamond detector was placed at P3 and the laser-driven proton bunch was focused onto it. The diamond detector signal was recorded by a fast oscilloscope and then deconvoluted to derive the spectrum of the transported beam \cite{Jahn2015}. Fig.~\ref{fig:1sol}d) compares the normalised spectrum with the simulation model prediction for $\alpha_3 = \SI[separate-uncertainty = true]{1.14+-0,05}{}$. The translation factor was found by minimising the deviations of the two datasets in accordance to cut-off energy and spectral shape.

As both solenoids are identical in construction, the same translation factor $\beta=\alpha$ can be used for S2 within the simulation model. To verify this assumption experimentally, in analogue to the TOF method, we compared an RCF stack measurement with simulated energy distributions of approximately \SI{25}{\mega\electronvolt} protons focused by the coupled system (S1 and S2) at position P4 (see fig.~\ref{fig:1sol}e)). By scanning the translation factors, the best agreement between model and measured spectrum (within \SI{10}{\percent}) was achieved for $\beta=\alpha=\SI{1.09}{}$. This value is perfectly consistent with the range provided by the $\alpha_{0} - \alpha_{3}$  measurements and therefore used as translation factor in all following simulations.  

\section*{Beamline optimisation for irradiation experiments at Draco} \label{sec:sec2}

In the following we present an experimental study on the optimisation of the beamline setup in particular for radiobiological irradiation studies at the \textit{Draco} laser facility. The previously verified beamline simulation model provided us with valuable input to achieve our ultimate goal -- the generation of complex dose distributions tailored to match a multitude of samples and applications by tuning the beamline parameters.  

One particular aim, which is further used as an example for the beamline optimisation, is the irradiation of a volumetric tumour on a mouse ear, according to Oppelt et al\cite{Oppelt2015,Beyreuther2017}. This tumour model was specifically designed to match the capabilities of a laser-driven proton beamline. The nearly spherical tumour has a diameter of approximately \SI{3}{\milli\metre}. A minimum proton range of \SI{5}{\milli\metre} in water was deemed necessary to account for tumour penetration including size and shape deviations as well as dosimetric control measurements in front of and behind the mouse ear, e.g. with RCFs. This penetration depth requires protons with a kinetic energy of at least \SI{25}{\mega\electronvolt}. Similarly, the diameter of the irradiation field was set to \SI{5}{\milli\metre}. An integrated dose of \SI{10}{\gray} has to be applied via multiple proton pulses within \SI{10}{\minute} to apply the necessary minimal net dose rate of \SI{1}{\gray\per\minute}\cite{Brenner1991}. The radiobiological model requires that every part of the volumetric tumour absorbs the identical proton dose. Hence, the lateral as well as the depth dose distribution have to be uniform, with an acceptable deviation of \SI{+-5}{\percent}. The acceptable dose deviation also applies for mean absorbed dose values throughout the pool of irradiated specimens.

\begin{figure}[htb]
\centering
\includegraphics[width=0.95\linewidth]{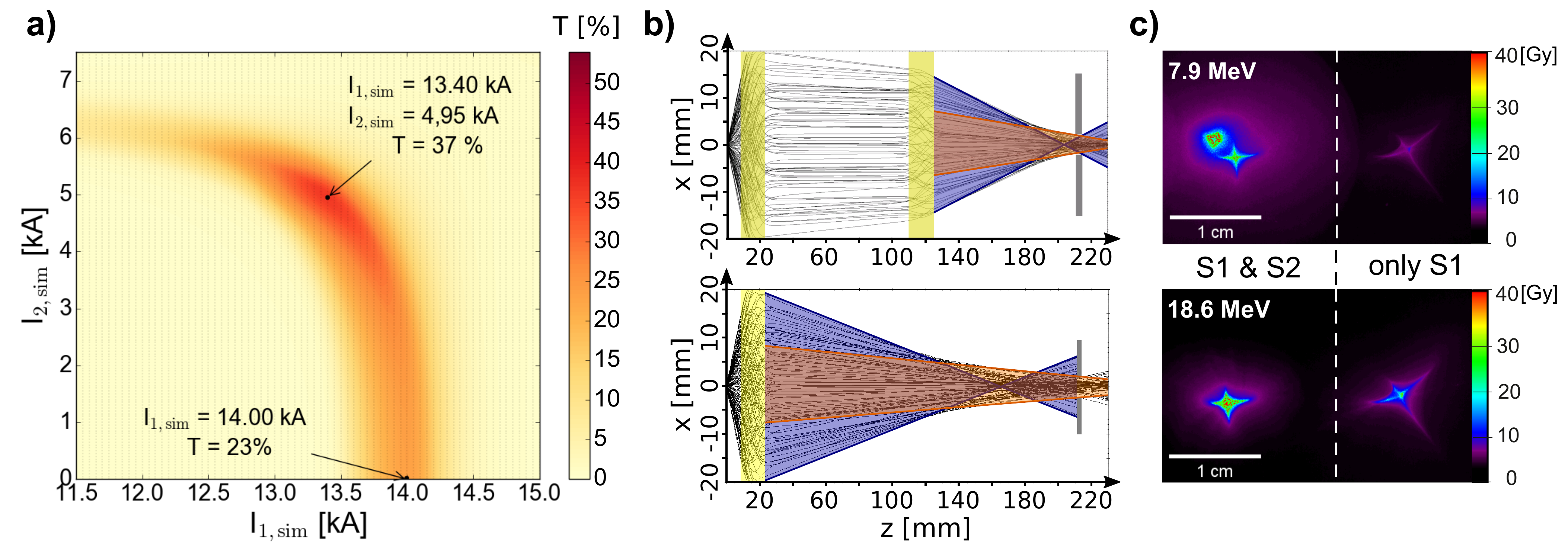}
\caption{\textbf{a)} Transmission efficiency (T) heat map for protons with \SI[separate-uncertainty = true]{25+-1}{\mega\electronvolt} as a function of solenoid currents $I_\mathrm{1,sim}$ and $I_\mathrm{2,sim}$. The lower indicated point marks the maximum transport efficiency for using only S1, the upper one the most efficient setting for dual solenoid transport. \textbf{b)} Associated proton trajectories from GPT simulation. The light yellow areas mark the solenoids. As guides for the eye, highly divergent incoming protons are marked in blue. They are focused closer to the source than protons with low initial divergence (orange). Dark grey boxes sketch the final aperture defining the irradiation area; only particles propagating through count as transmitted by the beamline. \textbf{c)} Lateral dose distribution for two chosen energies from one RCF stack at P4, dose-evaluated data from the films shown in fig.~\ref{fig:setup}c). Dual solenoid transport (left) yields two separate focal spot features of different kinetic energies, i.e  $\sim\SI{8}{\mega\electronvolt}$ and $\sim\SI{19}{\mega\electronvolt}$. The right focus was formed operating only solenoid S1.}
\label{fig:2sol}
\end{figure}

Taking into account the spectrum shown in fig.~\ref{fig:setup}b) and the required energy of \SI{25}{\mega\electronvolt} with a bandwidth of $\pm \SI{1}{\mega\electronvolt}$, approximately \SI{5.7E9}{} protons are generated and available for dose delivery. If all deposit their kinetic energy fully inside the tumour, the applicable dose would be \SI{23.3}{\gray}, exceeding the total dose requirement by a factor of two in a single pulse.
Fig.~\ref{fig:2sol}a) shows the theoretically predicted transmission efficiencies of the beamline for a proton beam with a spectral bandwidth of \SI[separate-uncertainty = true]{25+-1}{\mega\electronvolt} and TNSA-like divergence. To simplify the simulations, the protons are homogeneously distributed over a solid angle in accordance to the maximum divergence extracted from fig.~\ref{fig:setup}b), instead of the experimentally seen Gaussian-like distribution. Therefore, the simulation provides a lower limit of the transmission efficiency under optimised conditions. A parameter sweep over solenoid currents $I_\mathrm{1,sim}$ and $I_\mathrm{2,sim}$ was performed to find the operation point for maximum transmission. Protons are counted as transmitted when they hit a defined reference area of \SI{5}{\milli\metre} diameter at P5.
The $x$-axis in fig.~\ref{fig:2sol}a) represents single solenoid transport. The rest of the heat map area corresponds to the dual solenoid case. Indicated are the points of maximum transmission. Associated particle trajectories are shown in fig.~\ref{fig:2sol}b). 

Under optimised conditions, single solenoid transport provides a transmission efficiency of \SI{23}{\percent}. The first reason for particle loss is clipping of the highly divergent input beam at the capturing solenoid's aperture. The second reason is the spherical aberration of the solenoid lens. The highly divergent protons (depicted blue in fig.~\ref{fig:2sol}b), bottom) travel through the solenoid in close proximity to its windings, where the $B$-field is stronger. They are therefore focused closer to the solenoid and diverge afterwards to beam diameters larger than the reference area, so are not counted as transmitted.
For advanced spectral shaping, which will be explained below, and in order to enhance the beamline transmission we use solenoid S2 along with S1. With our simulation model a maximum transmission efficiency of \SI{37}{\percent} was predicted. As seen in the top picture of fig.~\ref{fig:2sol}b), under optimised conditions, the protons are almost collimated between the solenoids. Therefore, upon entering S2, no clipping occurs. The shorter focal length of S2 reduces the influence of spherical aberration on the overall transport. In the presented case, all protons entering solenoid S2 pass through the reference area. 

We transferred the optimised transmission conditions from GPT to experiment via translation factors $\alpha$ and $\beta$. 
The beamline model allows us to predict optimised transport parameters to match proton beams of various kinetic energies, practically enabled via the tuneable pulsed solenoids. The first comparative experimental studies on single and dual solenoid transport have been carried out at slightly reduced kinetic energy levels with the solenoid peak currents scaled accordingly. The beamline transmission has been empirically optimised beyond simulation model predictions by applying slight changes to the currents. Fig.~\ref{fig:2sol}c) shows associated lateral dose distributions from RCFs at focal position, i.e. P4, using both solenoids (left) and only the capturing solenoid S1 (right). 

We analysed data from three consecutive shots, comparing the number of transported protons at \SI{18.6}{\mega\electronvolt} with respective source characteristics and derived transmission efficiencies of \SI{50.6}{\percent} for dual solenoid and \SI{28.6}{\percent} for single solenoid transport, thus above theoretical prediction. We attribute this to the Gaussian angular distribution of TNSA protons (cf. fig.~\ref{fig:setup}b)) that was not featured in the simulation model. 
Comparing the numbers of transported protons for the consecutive measurements in fig.~\ref{fig:2sol}c), we derive an enhancement in transmitted particles by a factor of \SI{1.77}{} for the dual solenoid case over single solenoid. Comparing this to the respective ratio calculated from the simulation prediction, i.e. \SI{1.61}{}, we find very good agreement between theory and experiment.

\begin{figure}[tbp]
\centering
\includegraphics[width=0.95\linewidth]{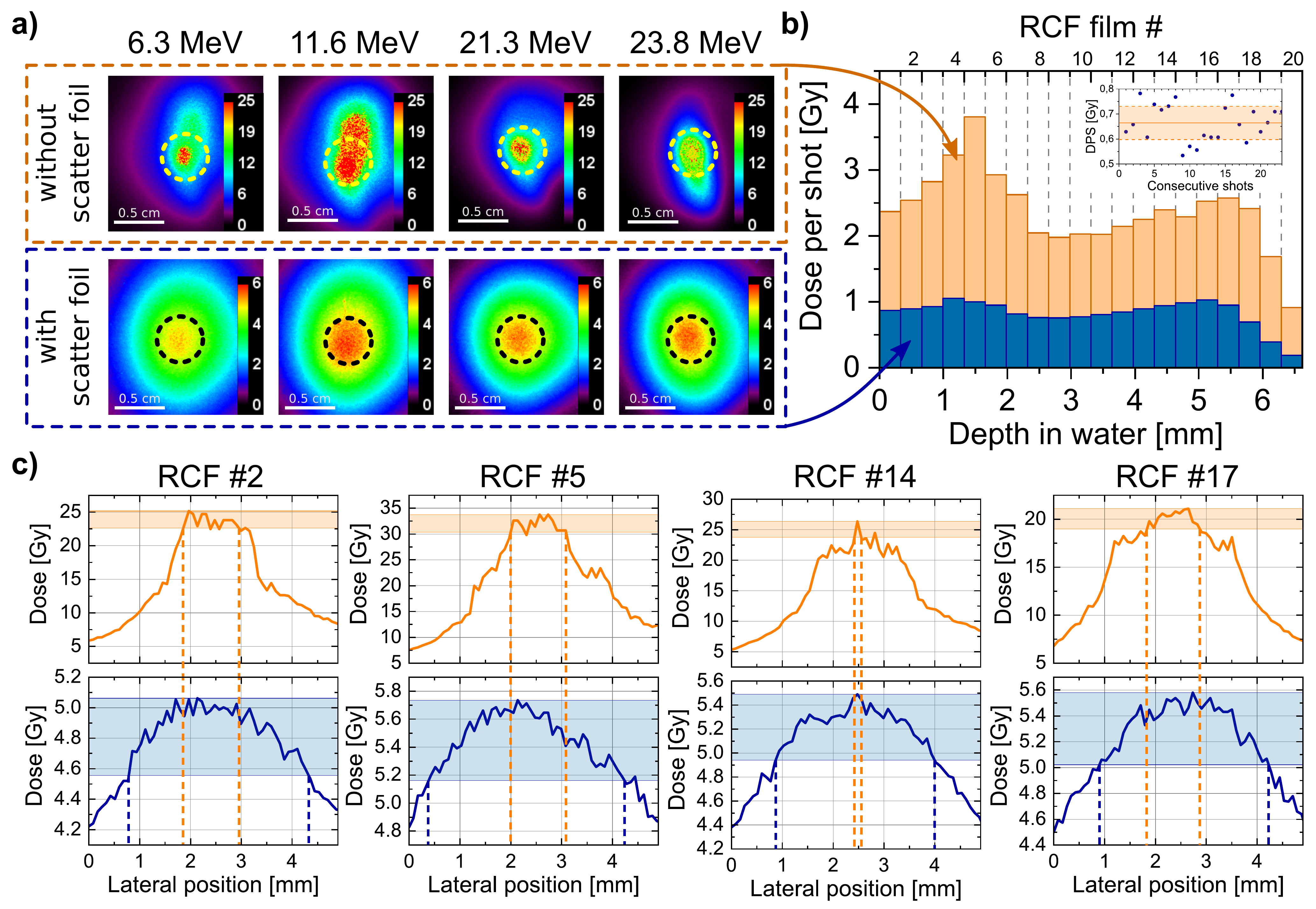}
\caption{Generation of homogeneous dose distributions via pulsed high-field beamline.
\textbf{a)} Compilation of RCF dose pictures (colour scale in \si{\gray}). Films at the top were irradiated at P5 without scatter foil and final aperture (\SI{5}{\mm} aperture size depicted in yellow). Two dose features corresponding to the two transported energy components can be distinguished. Films at the bottom show homogenised lateral dose distributions at P5 when a \SI{25}{\micro\metre} brass scatter foil is introduced at P4, a \SI{5}{\mm} aperture is depicted in black. The RCF data was acquired via cumulative irradiations with five consecutive proton pulses. \textbf{b)} Associated depth dose profiles. The mean dose values were evaluated over the area of the final aperture with \SI{5}{\milli\metre} diameter and are presented as mean dose per shot. Inset: Exemplary dose stability with indicated \SI{10}{\percent} variation interval shown as dose per shot (DPS) for 23 consecutive shots. \textbf{c)} Comparison of the lateral dose distribution of lineouts of \SI{1}{\mm} width across the \SI{5}{\milli\metre} aperture diameter. The RCF numbers and pictures correspond to the RCFs shown in a). The top row shows the case without scatter foil, bottom row shows the scattered case. The coloured area represents \SI{10}{\percent} peak to valley deviation in dose homogeneity. The ripples in the profiles originate from the limited spatial resolution of RCFs in combination with digitisation and are common in RCF data analysis.}
\label{fig:controlledirrad}
\end{figure}

Following up on the radiobiological scenario, fig.~\ref{fig:controlledirrad} summarises the first results of a proof-of-principle dual solenoid irradiation scheme, where high-resolution (absorber-free) RCF stacks were placed at the actual irradiation site P5, mimicking a radiobiological sample. 
The generated focal spots are currently too small and inhomogeneous to apply the dose to the sample, therefore we shifted the focal spot position closer to S2, letting the protons diverge after the focus to diffuse laterally.
As a result, higher energy protons are now focused at the sample position. Since they are undesired for our irradiation studies, these protons were suppressed by an energy selecting aperture of \SI{4}{\mm} diameter at P4.
Depicted in the top row of fig.~\ref{fig:controlledirrad}a) (five consecutive pulses integrated), the beam diameter is now large enough to cover the experimentally required \SI{5}{\milli\metre} irradiation field size. 
The lateral homogeneity was improved by introducing a scatter foil at P4 behind the energy selecting aperture.
Experimentally, \SI{25}{\micro\metre} brass was found to be best suited.
The main transported spectral component centred around \SI{25}{\MeV} exhibits a mean scatter angle of \SI{0.6}{\degree}, matching the geometrical half angle of \SI{0,5}{\degree}, which the final aperture spans between P4 and P5. The second spectral component around \SI{12}{\MeV} is scattered more strongly, i.e. \SI{1.2}{\degree}. An RCF stack was irradiated with five consecutive transported proton pulses in this beamline configuration. Corresponding colour-coded dose pictures are presented in fig.~\ref{fig:controlledirrad}a), bottom row, where the black circle represents the aperture size of \SI{5}{\milli\metre}.
Fig.~\ref{fig:controlledirrad}c) shows lineouts of the lateral dose homogeneity along the aperture diameter averaged over \SI{1}{\milli\metre}. 
The envisaged peak to valley deviation of \SI{10}{\percent} is indicated by the shaded area. Comparing the lateral homogeneity between unscattered (orange dashed lines) and scattered case (blue dashed lines), significant improvement is visible for all depths. For the latter case, a homogeneous dose distribution over a circle of \SI{3}{\milli\metre} was achieved which corresponds to the envisaged tumour size in the dedicated mouse model.

For the presented RCF stacks we calculated the mean dose value of each film within the \SI{5}{\milli\metre} irradiation area divided by the number of shots and plotted these values in form of a depth dose distribution to show the mean dose per shot (see fig.~\ref{fig:controlledirrad}b)). Depicted in orange is the depth dose distribution for dual solenoid transport without scatter foil. We see remaining inhomogeneities throughout the stack due to the intense low-energy component of the transported protons in accordance to TNSA source characteristics. Introducing the scatter foil allows us to generate a homogeneous dose distribution in depth. The dose in shallow depths is decreased above average, because of the stronger scattering that the low energy protons undergo when passing the brass scatter foil. Therefore they are spread over a larger area at the sample position. 

For proton source characteristics close to those in fig.~\ref{fig:setup}b), the beamline enabled single shot mean dose values of around \SI{2.3}{\gray}. Upon introducing the scatter foil the mean dose per shot was determined to be around \SI{0,7}{\gray}. 
This trade-off, however, allows a homogeneous dose delivery to a cylindrical volume of \SI{5}{\mm} diameter and \SI{5}{\mm} depth with a mean depth dose homogeneity of \SI{+-8.5}{\percent} and the lateral dose homogeneity depicted in fig.~\ref{fig:controlledirrad}c), already very close to the radiobiologically required maximum dose deviation of \SI{+-5}{\percent} over the whole volume.
To reach the required dose rate of \SI{1}{\gray\per\minute}, about three pulses have to be applied within two minutes. This is presently feasible by the experimental setup since it enables two pulses per minute. The beamline has proven to work reliably over $\sim 1000$ shots. The inset of fig.~\ref{fig:controlledirrad}b) shows exemplary data on dose per shot at P5 from 23 consecutive shots, derived from the charge generated by the protons passing through the transmission ionisation chamber. 

In summary, a multitude of spectral shapes and therefore depth dose distributions can be generated at the sample position. Furthermore, adjusting the solenoids' lateral position and orientation with respect to each other leads to a spatial separation of the low- and high-energy focus and can therefore be used to fine-tune the ratio of both spectral components, e.g. to counter changes in the slope of the exponential source spectrum. By altering the distance between S1 and S2 as well as field strengths, complementary to changing the intensity of the transported energy components, we can tune their spectral separation. 
The possibility of introducing various apertures and scatter bodies on demand at multiple positions along the beamline extends the capabilities of the beam transport system even further.

\section*{Conclusion and Outlook}

We set up and experimentally optimised a pulsed high-field magnet beamline for laser-driven protons at the \textit{Draco} PW laser.
The beamline uses two solenoid lenses to transport broader parts of the TNSA source spectrum relative to the quasi-monoenergetic transport via a single solenoid lens. As a result, a single-shot spread-out Bragg-Peak can be generated, otherwise only accomplished by introducing complicated ridge filters\cite{Nakagawa2000} or by applying the dose via multiple shots. In this manner, the beamline delivers a homogeneous depth dose distribution,  required for radiobiological irradiation studies. 

Optimising for a dedicated mouse tumour model, which demands a homogeneous dose distribution with \SI{+-5}{\percent} deviation over a cylindrical volume of \SI{5}{\milli\metre} depth and diameter, an adapted beamline setup was established. We achieved a depth dose homogeneity of \SI{+-8.5}{\percent}.
Laterally, a dose homogeneity of \SI{+-5}{\percent} was achieved for a circle of at least \SI{3}{\milli\metre} diameter throughout the entire target volume.  For this setup a dose per shot of \SI{0.7}{\gray} was demonstrated, allowing for a dose rate of more than \SI{1}{\gray\per\minute}. In conclusion, we established a controlled transported dose distribution close to the model requirements, keeping the ambitious goal of a radiobiologically relevant volumetric tumor irradiation in reach.

\begin{figure}[tbp]
\centering
\includegraphics[width=0.95\linewidth]{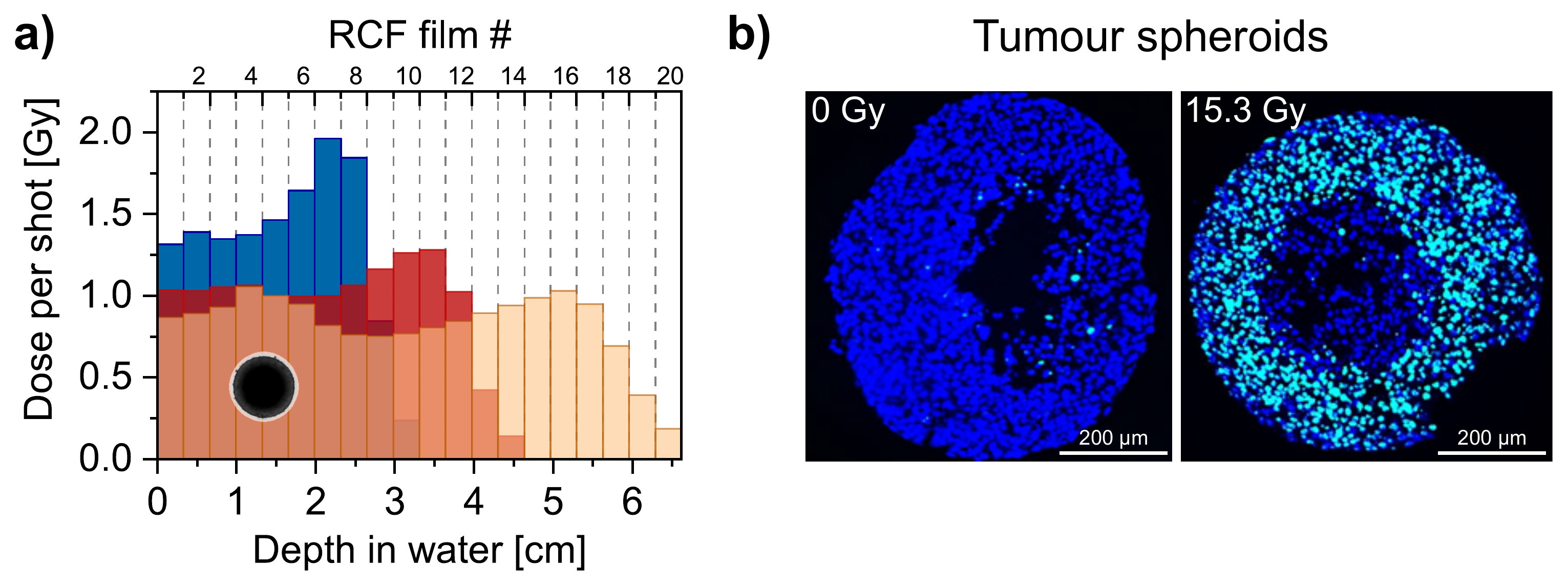}
\caption{\textbf{a)} Depth dose distributions (RCF measurement) for different irradiation setups (solenoid currents) while keeping the input spectrum constant.
\textbf{b)} Median sections (\SI{10}{\micro\metre} thickness) through an unirradiated (left) and a \SI{15.3}{\gray} irradiated tumour spheroid (right) both labelled for DNA double strand breaks (bright spots). After irradiation, a ring of laser-driven proton induced DNA DSBs is clearly visible. The spheroid centre does not show DNA DSBs because of a necrotic area, common to spheroids of this size (see methods section for details).}
\label{fig:irrad}
\end{figure}

In order to experimentally demonstrate the tuneability and in view of future radiobiological studies, fig.~\ref{fig:irrad}a) shows three different transported depth dose distributions adjusted with our proton beamline. Recognisably, the dose per shot increases, 
when decreasing the transported energy to adapt for smaller sample sizes, while keeping all other conditions unchanged. We attribute this to the exponentially decaying laser-driven proton source spectrum and conclude that studies with smaller irradiation targets (e.g. tumour spheroids, schematically shown in fig.~\ref{fig:irrad}a)) may be conducted with larger pulse dose rates if required.

We chose to perform dose-controlled irradiations of \textit{in-vitro} tumour spheroids (volumetric cluster of human squamous cell carcinoma of the tongue (SAS cell line), $\sim\SI{650}{\micro\metre}$ diameter, scaled sketch in fig.~\ref{fig:irrad}a)) as technical proof-of-concept scenario. We placed our focus not on the generation of a radiobiologically relevant outcome but on the interplay between all procedures and experimental challenges necessary for a future radiobiology campaign, in particular (1) biological sample handling and beamline implementation, (2) laser-driven proton acceleration, (3) beam shaping according to the radiobiological specifications, (4) online and offline dosimetry and (5) analysis of radiation induced effects. 
The samples were irradiated with \SI{15.3}{\gray}\SI{+-15}{\percent} (uncertainty due to homogenisation ca. \SI{7}{\percent} and dosimetry with the combination of RCF and IC ca. \SI{8}{\percent}) using the red depth dose profile depicted in fig.~\ref{fig:irrad}a). 
As the result, clear signatures of acute DNA double strand breaks (DSB) are visible by evaluating $\gamma$-H2AX foci as surrogate markers in the irradiated sample as opposed to unirradiated controls (cf. fig.~\ref{fig:irrad}b)). While this irradiation is clearly not a comprehensive radiobiological study, it marks an important step towards full scale studies in the near future.

Looking beyond the scope of our specific goals, recent works show a growing interest in radiobiological effects of high mean dose rate radiation\cite{Hanton2019,Wilson2020}.
From the presented results we expect to provide an attractive research platform to radiobiologists working on said topic in the future. Therefore, and to enable experiments of even larger scale, we are further improving the beamline:
The first pursued approach is increasing the pulse repetition rate, to be accomplished via cooled high-field solenoid magnets powered by high-repetition rate current pulse drivers\cite{Wettengel2018}. 
Ultimately, beamline operation rates of \SI{1}{\hertz} are the goal and currently under development, to match the repetition rate of \textit{Draco} PW and advanced targetry\cite{Gauthier2017}.
The second approach relies on improving the source spectra to provide more protons of desired energy to the beamline. 
Recent proton energy scaling studies performed with the same laser system suggest an enhancement in particle flux around \SI{25}{\MeV} by about one order of magnitude when proton beams with cut-off energies in the range of \SI{50}{\MeV} are generated in the TNSA mechanism.
Our current beamline setup is designed for maximum proton energies of up to \SI{70}{\mega\electronvolt}, 
enabling bigger sample sizes and more complex radiobiological scenarios in future studies.

\section*{Methods} \label{sec:methods}

\subsection*{Solenoids, magnetic field characteristics and pulsed current drivers} \label{subsec:p+beamline}
Two identically constructed solenoid coils have been installed \SI{8}{\centi\meter} (first winding of solenoid S1) and \SI{110}{\centi\meter} (first winding of solenoid S2) behind the laser target. The first solenoid acts as a capturing device for the highly divergent protons to either focus them onto a sample/detector directly or transport the collimated beam towards the second solenoid which in turn focuses. Due to their distance, an influence of the magnetic field of S1 to S2 can be neglected (or vice versa).
For remote controlled alignment, solenoid S1 is mounted on a hexapod (Newport HXP100-MECA). Solenoid S2 is mounted on a table with height adjustable screws, which itself is reproducibly movable with the same degrees of freedom as S1, but not motorised. The solenoids were manufactured at the Dresden High Magnetic Field Laboratory (HLD). They consist of 112 evenly distributed windings in four layers, each reinforced by at least \SI{1.5}{\milli\meter} Zylon. A strong copper alloy (Wieland K88) wire of \SI{4.3 x 2.8}{\milli\meter} cross section is wound on a cylindrical fibre-reinforced plastic (FRP) body with \SI{54}{\mm} outer diameter and \SI{48}{mm} inner bore size. This large bore size allows for a high capture efficiency. The capturing solenoid S1 is enclosed by a cylindrical stainless steel housing with a PEEK lid. The housing is connected via bellows to the outside of the target chamber to operate the coil in ambient air. This measure has to be taken to prevent the reinforcement and FRP from outgassing, which would deteriorate the vacuum quality and lower the electrical insulation of the coil. The housing also shields the coil from direct exposure to the high flux ionising radiation.

Using the beamline for proton beam transport, the \si{\nano\second} long proton pulse is sent through the solenoid, when the pulsed $B$-field is at the maximum of its \si{\milli\second} long pulse and appears therefore stationary for the protons (see fig.~5 in Zeil et al\cite{Zeil2013a}). Although the magnetic field appears in a ``stationary'' regime when the protons pass through, field distortions, additional fields induced by eddy currents or other asymmetric field influences induced by the current pulse have to be taken into account. They can be several orders of magnitude lower than the peak intensity (of the pulsed B-field) and are difficult to measure, as they are superimposed with the high peak field. These errors do not (necessarily) appear in DC mode and additionally to the missing field deviations, measuring in DC mode decreases the peak field intensity by orders of magnitude (around a factor of 1000). The (orders of magnitudes lower) deviations in field strengths are then even harder to detect, which depicts a DC mapping for comparison of the pulsed field unsuitable. 3D field mapping in a pulsed mode either with a suitable Hall probe or with other common detectors for accelerator magnets (e.g. pick-up coils, pulsed wires or rotating coils) remains impractical due to said detection issues and the tremendous amount of pulses necessary for a reasonably dense field map. This leads to the semi-empiric approach of the beamline characterisation via proton transport presented in the section ``Beamline modelling and experimental verification''. 

To provide the coils with high current pulses, two capacitor-based pulse generators ($C_{\mathrm{total}} = \SI{326}{\micro\farad}$ and $C_{\mathrm{total}} = \SI{200}{\micro\farad}$) have been developed which can provide maximum charging voltages of $U_{\mathrm{C}} = \SI{-24}{\kilo\volt}$ and $U_{\mathrm{C}} = \SI{-16}{\kilo\volt}$ resulting in stored energies of \SI{94}{\kilo\joule} and \SI{25,6}{\kilo\joule} respectively. Capacitor bank and solenoid (inductance $L \approx \SI{270}{\micro\henry}$) together act as a resonant circuit where the capacitor is discharged over a high-voltage switch (thyratron) into the high-field solenoid, yielding an on-axis magnetic field strength of up to $B_{\mathrm{max}} \approx \SI{19,5}{\tesla}$. The $B$-field strength is not directly measured in the experiment but can be derived from online pulse current measurements via Rogowski coil (Power Electronic Measurements Ltd, CWT 1500, measurement error $<\SI{+-1}{\percent}$) recorded for every pulse.

The maximum repetition rate of the pulse generators is around three pulses per minute, but applying the highest current to the solenoids over a time of about more than ten minutes leads to a temperature increase due to Ohmic heating in the solenoid, which endangers the compound of the reinforcement and adhesive. This could lead to outgassing and lowering of the insulation, enabling a possible spark between the windings and therefore has to be avoided. This repetition rate in combination with the repetition rate of the laser-proton source (mostly limited by target alignment) results in an effective repetition rate of two pulses per minute. 

\subsection*{GPT-Simulation}

For the simulations we reproduced the solenoids in "General Particle Tracer" in a simplified way, by modelling 112 current loops with distances according to the actual manufactured solenoid. The $B$-field is computed by GPT via the input current specified by the user. The proton source resembles a TNSA source with minor restrictions. It is implemented as a point source, the energy dependent proton number follows the spectral shape shown in fig.~\ref{fig:setup}b), top, while the energy dependent divergence follows fig.~\ref{fig:setup}b), bottom. Within one energy bin (\SI{100}{\keV}), the particles are distributed homogeneously over the respective divergence angle. For particle tracing, GPT calculates the  three-dimensional relativistic equations of motion of particles as function of time. The solver uses a fifth-order Runge-Kutta method with adaptive step size control.\cite{Loos1996} 

\subsection*{Diagnostics and irradiation site} \label{subsec:diagnostics}

A scintillator detector block (BC-408, Saint-Gobain Crystals, \SI{1}{\cm} thickness) was used at P4 or P5 (see fig.~\ref{fig:setup}) for beamline alignment purposes and as an online monitor for an approximation of the lateral and depth dose distribution of the transported proton beam.

For more precise energy and dose resolving measurements (offline), stacks of self-developing radiochromic films (RCF) of type Gafchromic EBT3 replace the scintillator. RCFs are sensitive to ionising radiation and darken correspondingly to the dose they have been exposed to. With an appropriate calibration, a single film provides two-dimensional (lateral) dose information. When used as a stack of several films, they allow to map three-dimensional dose distributions and, via deconvolution, to reconstruct the kinetic energy spectrum of the impinging proton beam\cite{Nurnberg2009}. Our RCFs were calibrated at an X-ray source at our facility and then scanned using a calibrated flatbed scanner. The calibration data was cross checked with a calibration of identical EBT3 films (same batch) performed using a clinical proton source.

To reach the in-air irradiation site, particles exit the diagnostic chamber through a calibrated transmission ionisation chamber (IC, PTW X-Ray Therapy Monitor Chamber 7862). 
The employed IC has been cross calibrated using RCF stacks at sample position. As a result, changing the beamline transport setting requires a new calibration. For all presented setups these calibration procedures have been thoroughly conducted. 
Therefore, the transmission IC is used as a tool to predict the dose delivered to the sample, which was already proven to be reliable for radiobiological studies with laser-driven protons\cite{Richter2011}.

Protons propagating further to the sample traverse through up to \SI{30}{\centi\metre} of air. The resulting scattering of the proton beam may be taken into account for irradiation studies, while the energy loss of the particle is negligible (well below \SI{1}{\percent}). 

Irradiation samples are positioned behind a stainless steal aperture of \SI{5}{\mm} diameter to precisely control the irradiated area. The aperture thickness was chosen to be \SI{5}{\mm}, thick enough to block protons of up to \SI{55}{\MeV} kinetic energy.

\subsection*{Biological tissue handling, irradiation and analysis} \label{subsec:bio}

The Tumour spheroids were grown in liquid overlay according to the protocol by Friedrich et al\cite{Friedrich2009}. Here, we seeded 7000 cells from the cell line SAS (RRID: CVCL\_1675, human squamous cell carcinoma of the tongue) per well into agarose-coated 96-well plates and used the spheroids at day 6 in culture with a diameter of \SIrange{600}{650}{\micro\metre}. This spheroid size is generally associated with proliferation gradients and the presence of hypoxia and central secondary necrosis reflecting the pathophysiology in tumour microregions. For the treatment, single spheroids were transferred into a cuvette filled with \SI{50}{\micro\liter} agarose and a nutrient solution (DMEM, penicillin/streptomycin (\SI{1}{\percent}),  fetal calf serum (\SI{20}{\percent})) and then irradiated. 
For the irradiation one cuvette was placed at irradiation side; as control sample three other ones were stored next to beamline, but protected from radiation background. In addition to the controls, that move along the treated samples, some spheroids remain in the lab under standard conditions in order to reveal potential influences from the laser environment. The control sample was treated similarly except irradiation.

The irradiations at the \textit{Draco} PW have been performed using the dual solenoid setup with \SI{25}{\micro\metre} brass scatter foil, again resulting in a laterally homogenised dose distribution and with maximum deviations of below \SI{7}{\percent} up to a depth of \SI{2.5}{\milli\metre} (red setting shown in fig.~\ref{fig:irrad}a)). This dose profile was chosen by taking into account the spheroids' size, thickness of the cuvette, nutrient and spheroid medium as well as RCFs in front and back for quality assurance.

After treatment, all cuvettes were transported back into the lab, and the spheroids were transferred into primed 96 well plates containing spheroid medium and \SI{20}{\micro\gram\per\milli\liter} Pimonidazole (hypoxia marker pimonidazole (Hydroxyprobe Omni Kit, Natural Pharmacia Int., Burlington, MA, USA), and maintained under standard conditions (\SI{37}{\celsius}, \SI{8}{\percent} CO$_2$) for two hours. Subsequently, they were fixed in \SI{4}{\percent} paraformaldehyde for 24 hours to be embedded in paraffin for sectioning and further analysis. The \SI{10}{\micro\metre} median sections were imaged using an Axiovert S100 with a magnification of 100. 

Cell nuclei (in detail the DNA) and hypoxic areas were stained with DAPI (4-6-Diamidino-2-phenylindole, Axxora, Lörrach, Germany) and an antibody against pimonidazole according to manufacturer’s instructions. DNA double-strand breaks were marked by $\gamma$-H2AX staining according to Beyreuther et al\cite{Beyreuther2009}. $\gamma$-H2AX requires a living cell that is able to start the radiation damage repair process by activating the histone H2AX via phosphorylation. Hence, the staining did not only show the DNA (DSBs), but also the vital part of the spheroid delimiting it from the dead, necrotic central area (fig.~\ref{fig:irrad}b), right). For comparison, an exemplary control spheroid is shown on the left exhibiting just a few background DSB. The larger black area/hole in the middle could also be attributed to an artefact of the cutting of the paraffin-embedded spheroid. 

\bibliography{references}

\section*{Acknowledgements}

S Bock, R Gebhardt, U Helbig and T Püschel are highly acknowledged for their laser support. We thank T Hermannsdörfer, S Zherlitsyn and the workshop of the High-field Laboratory Dresden (HLD) for their advice and the manufacturing of the solenoids. We thank NJ Hartley for his help preparing the paper.
We also thank M Wondrak for her technical assistance with spheroid preparation. The work was partially supported by EC Horizon 2020 LASERLAB-EUROPE/LEPP (Contract No. 654148 and 871124). 

\section*{Author contributions statement}

F.-E.B., F.K., C.B., L.G., L.K., S.K., J.M.-N., L.O.-H., J.P., M.R., H.-P.S., T.Z. and K.Z. prepared and conducted the experiment. E.B., E.L., L.K.S., E.R.S. handled and evaluated the biological samples. F.-E.B. and F.K. analysed the results. M.S. supervised the fabrication of the solenoids. T.E.C., U.S. and K.Z. supervised the project. All authors contributed to discussions and revision of the manuscript.

\section*{Additional information}
 
\textbf{Competing interests:} The authors declare that they have no competing interests.

\end{document}